\documentclass[aps,showpacs,preprintnumbers,amsmath,amssymb]{revtex4}

\usepackage{graphics,epsfig}
\usepackage{graphicx}
\usepackage{dcolumn}
\usepackage{bm}

\newcommand{\be}{\begin{equation}}
\newcommand{\ee}{\end{equation}}
\newcommand{\ba}{\begin{eqnarray}}
\newcommand{\ea}{\end{eqnarray}}

\begin{document}

\baselineskip=0.6cm
\title{Asymptotic tails of massive scalar fields
in a stationary axisymmetric EMDA black hole geometry}
\author{Qiyuan Pan and Jiliang Jing} \email{Corresponding author, jljing@hunnu.edu.cn}
\affiliation{ Institute of Physics and  Department of Physics, \\
Hunan Normal University,\\ Changsha, Hunan 410081, P. R. China }

 \begin{abstract}
The late-time tail behavior of massive scalar fields is studied
analytically in a stationary axisymmetric EMDA black hole
geometry. It is shown that the asymptotic behavior of massive
perturbations is dominated by the oscillatory inverse power-law
decaying tail $\ t^{-(l+3/2)}\sin(\mu t)$ at the intermediate late
times, and by the asymptotic tail $\ t^{-5/6}\sin(\mu t)$  at
asymptotically late times.  Our result seems to suggest that the
intermediate tails $ t^{-(l+3/2)}\sin(\mu t)$ and the
asymptotically tails $t^{-5/6} \sin(\mu t)$ may be quite general
features for evolution of massive scalar fields in any four
dimensional asymptotically flat rotating black hole backgrounds.
\vspace{0.2cm}

\end{abstract}

\vspace*{0.4cm}
 \pacs{03.65.Pm, 04.30.Nk, 04.70.Bw, 97.60.Lf}

\maketitle

 The no-hair theorem, introduced by Wheeler in the early 1970s
\cite{1,2}, states that the external field of a black hole relaxes
to a Kerr-Newman field characterized
 solely by the black-hole's mass, charge and angular
momentum. So it is interesting to analyze
the dynamical mechanism by which perturbation fields
outside a black hole are radiated away. Price first
studied the massless neutral external perturbations
in \cite{3} and found that the late-time behavior
is dominated by the factor ~$t^{-(2l+3)}$
for each multiple moment ~$l$. Hod and Piran investigated
the late-time evolution of a charged massless scalar field in  Refs.
\cite{4}-\cite{6} and their conclusion was that
a charged scalar hair outside a charged black hole is
dominated by a ~$t^{-(2l+2)}$ tail which
decay slower than a neutral one. The asymptotic late-time
tail of massless perturbations outside realistic, rotating
black hole had been considered in Refs. \cite{7}-\cite{9}. And recently
the massless Dirac perturbations have been examined in  Refs.
\cite{10}-\cite{12}.

Although these works are mainly concerned with massless fields,
the evolution of massive scalar hair is also important and it has
attracted a lot of attention recently. The physical mechanism by
which late-time tails of massive scalar fields are generated may
be qualitatively different from that of massless ones. It has been
shown \cite{5} that if the field mass ~$\mu$ is small, namely
~$\mu M\ll 1 $, the oscillatory inverse power-law behavior
~$t^{-(l+3/2)}\sin(\mu t)$, dominates as the intermediate
late-time tails. Recently the very late-time tails of the massive
fields in the Schwarzschild or Reissner-Nordstr\"{o}m background
were studied in Refs. \cite{14,15}, and it has been pointed out
that the oscillatory inverse power-law behavior of the dominant
asymptotic tail is approximately given by ~$t^{-5/6}\sin(\mu t)$
which is slower than the intermediate ones. The late-time tails of
a massive scalar field in the spacetime of black holes are studied
numerically by Khanna \cite{Khanna}.  Yu \cite{Hongwei} studied
the decay of massive scalar hair in the background of a black hole
with a global monopole.

Nowadays, it seems that the superstring theories are the most
promising candidates for a consistent quantum theory of
gravitation. So there have been many investigations concerning the
spacetimes of dilaton black hole. For instance, the late-time
tails of massive scalar fields in the background of dilaton black
hole had been explored by Moderski and Rogatko, and it had been
shown that this late-time evolution is identical with that in the
Schwarzschild and Reissner-Nordstr\"{o}m backgrounds \cite{16}.
Obviously they only discussed the massive power-law tails in the
static, spherically symmetric black hole geometry. Thus, it is
worthwhile to investigate analytically the massive late-time
behavior in the background of dilaton black hole being the
stationary axisymmetric solution of the so-called low-energy
string theory \cite{17,18}.

Now we begin to study analytically the asymptotic tails of massive
scalar fields in the background of a stationary axisymmetric
EMDA black hole. The metric for this considered black hole can be expressed
as \cite{17,18}
\begin{eqnarray}
\label{1}&&ds^{2}=-\frac{\Sigma-a^{2}\sin^{2}\theta}
{\Delta}dt^{2}-\frac{2a\sin^{2}\theta}{\Delta}
\left[\Xi -\Sigma\right]dtd\varphi+
\frac{\Delta}{\Sigma}dr^{2}\nonumber\\
&&\qquad~~+\Delta d\theta^{2}+\frac{\sin^{2}\theta}
{\Delta}\left[\Xi ^2-\Sigma a^{2}\sin^{2}
\theta\right]d\varphi^{2},
\end{eqnarray}
with ~$\Sigma=r^{2}-2Mr+a^{2},\Delta=r^{2}-2Dr+a^{2}
\cos^{2}\theta$, $\Xi =r^{2}-2Dr+a^{2}$,  and  $D$,  $a$, and $M$
represent the dilaton, rotational, and  mass parameter.

In the background geometry,  let massive scalar field as
$\Phi=\Xi ^{-1/2}
\sum_{m=-\infty}^{+\infty}\Psi^{m}e^{im\varphi}$,
one obtains a wave-equation for each value of  azimuthal number $m$:
\begin{eqnarray}
\label{3}&&C_{1}(r,\theta)\frac{\partial^{2}\Psi}
{\partial t^{2}}+C_{2}(r,\theta)\frac{\partial\Psi}
{\partial t }
+C_{3}(r,\theta)\Psi-\frac{\partial^{2}\Psi}
{\partial y^{2}}-\frac{\Sigma}{\Xi ^2}\frac{1}
{\sin\theta}\frac{\partial}{\partial\theta}
\left(\sin\theta\frac{\partial\Psi}{\partial\theta }
\right)=0,
\end{eqnarray}
where we use the tortoise radial coordinate ~$y$ defined by
~$dy=\Xi dr/\Sigma$, and $C_{i}(r,\theta)$ are given by
\begin{eqnarray}
\label{4}&&C_{1}(r,\theta)=1-\frac{\Sigma a^{2}\sin^{2}
\theta}{ \Xi ^2}, \nonumber  \\
\label{5}&&C_{2}(r,\theta)=\frac{4iam(M-D)r}{\Xi ^2},\nonumber  \\
\label{6}&&C_{3}(r,\theta)=\frac{\Sigma}{\Xi ^2}
\left[-m^{2}\left(\frac{a^{2}}{\Sigma}-\frac{1}{\sin^{2}
\theta}\right)
 +\frac{2(r-M)(r-D)+\Sigma}{r^{2}-2Dr+a^{2}}-\frac{3(r-D)^{2}\Sigma}{\Xi ^2}+
 \Delta\mu^{2}\right], \nonumber
\end{eqnarray}
where ~$\mu$ is the mass of the scalar field.

The time evolution of a wave field described by Eq. (\ref{3}) is
given by
\begin{eqnarray}
\label{7}&&\Psi(x,t)=2\pi\int\int_0 ^\pi \{C_{1}(x')
[G(x,x';t)\Psi_{t}(x',0)+G_{t}(x,x';t)\Psi(x',0)]
\nonumber\\
&&\qquad\qquad~+C_{2}(x')G(x,x';t)\Psi(x',0)\} \sin
\theta'd\theta'dy',
\end{eqnarray}
for ~$t>0$, where ~$x$ stands for ~$(y,\theta)$.
The Green's function ~$G(x,x';t)$ is defined by
\begin{eqnarray}
\label{8}&&\left[C_{1}(r,\theta)\frac{\partial^{2}}
{\partial t^{2}}+C_{2}(r,\theta)\frac{\partial}{\partial t }
+C_{3}(r,\theta)-\frac{\partial^{2}}{\partial y^{2}}
\right.\nonumber\\
&&\left.-\frac{\Sigma}{\Xi ^2}\frac{1}
{\sin\theta}\frac{\partial}{\partial\theta}
\left(\sin\theta\frac{\partial}{\partial\theta }\right)
\right]G(x,x';t)
 =\delta(t)\delta(y-y')\frac{\delta(\theta-\theta')}
 {2\pi\sin\theta}.
\end{eqnarray}
The causality condition gives us the initial condition
~$G(x,x';t)=0$ for ~$t\leq0$. We express the Green's function
in terms of the Fourier transform ~$\tilde{G}_{l}(y,y';\omega)$
\begin{eqnarray}
\label{9}\ \ \ \ G(x,x';t)=\frac{1}{(2\pi)^{2}}\sum_{l}
\int_{-\infty+ic}^{\infty+ic}\tilde{G}_{l}(y,y';\omega)
Y^{m}_{l}(\theta,a\omega)Y^{m}_{l}(\theta',a\omega)e^{-i
\omega t}d\omega,
\end{eqnarray}
where ~$c$ is some positive constant and ~$Y^{m}_{l}(\theta,a\omega)$
are the spheroidal harmonics.

The  Fourier transform  is analytic in the upper half
~$\omega-$plane and it satisfies
 \begin{eqnarray}
\label{11}&&\left[\frac{d^{2}}{dy^{2}}+\omega^{2}-R_{l}
(r,\omega)\right]\tilde{G}_{l}(y,y';\omega)=\delta(y-y'),
\end{eqnarray}
where
\begin{eqnarray}
\label{12}&&R_{l}(r,\omega)=\frac{\Sigma}{\Xi ^2}\left[a^{2}\omega^{2}+\frac{4am\omega (M-D)r}{\Sigma}
+l(l+1)+\mu^{2}(r^{2}-2Dr)\right.\nonumber\\&&\qquad\qquad~
\left.-\frac{m^{2}a^{2}}{\Sigma}+\frac{2(r-M)(r-D)+
\Sigma}{\Xi  }
-\frac{3(r-D)^{2}\Sigma}{\Xi ^2}\right].
\end{eqnarray}
We define two auxiliary functions ~$\tilde{\Psi}_{1}
(y,\omega)$ and ~$\tilde{\Psi}_{2}(y,\omega)$ which are
(linearly independent) solutions to the homogeneous
equation
\begin{eqnarray}
\label{13}&&\left[\frac{d^{2}}{dy^{2}}+\omega^{2}-R_{l}
(r,\omega)\right]\tilde{\Psi}_{i}(y,\omega)=0,\ \ i =1,2.
\end{eqnarray}
Then, the Green's function can be given by
\begin{eqnarray}
\label{14}\ \ \ \ \tilde{G}_{l}(y,y';\omega)=-\frac{1}
{W(\omega)}
\left\{ \begin{array}{l}
\tilde{\Psi}_{1}(y,\omega)\tilde{\Psi}_{2}(y',
\omega),\ \ y<y'; \\
\tilde{\Psi}_{1}(y',\omega)\tilde{\Psi}_{2}(y,\omega),
\ \ y>y'. \end{array} \right.
\end{eqnarray}
Here ~$W(\omega)=W(\tilde{\Psi}_{1},\tilde{\Psi}_{2})
=\tilde{\Psi}_{1}\tilde{\Psi}_{2,y}-\tilde{\Psi}_{2}
\tilde{\Psi}_{1,y}$ is the Wronskian.

To calculate ~$G(x,x';t)$ using Eq. (\ref{9}), we can
close the contour of integration into
the lower half of the complex frequency plane.
It has been argued that the asymptotic massive tail is
associated with the existence of a branch cut (in
~$\tilde{\Psi}_{2}$) placed along the interval
~$-\mu\leq\omega\leq\mu$ \cite{5,19}. As will be shown in this
Letter, this tail arises from the integral of
~$ \tilde{G}(y,y';\omega)$ around the branch cut
(denoted by ~$G^{c}$) which gives rise to an oscillatory
inverse power-law behavior of the field. So our goal
is to evaluate ~$G^{c}(x,x';t)$.

\ \ Assume that both the observer and the
initial data are situated far away from the black hole
so that ~$r\gg M-D$, we may expand the wave-equation (\ref{13})
as a power series in ~$\frac{M}{r}$, neglecting terms of
order ~$O(\frac{\alpha}{r^{n\geq2}})$. And if we
define ~$\varpi=\sqrt{\mu^{2}-\omega^{2}}$ and
~$z=2\varpi r$, we obtain
\begin{eqnarray}
\label{15}&&\left\{\frac{d^{2}}{dz^{2}}+\left[-\frac{1}{4}+
\frac{\beta}{z}+\frac{1/4-(l+1/2)^{2}}{z^{2}}\right]\right\}
\xi (r,\omega)=0,
\end{eqnarray}
where ~$\xi  (r,\omega)=[\Sigma/\Xi  ]^{1/2}
\tilde{\Psi}(y,\omega)$ and ~$\beta=(M-D)\mu^{2}/\varpi-2(M-D)\varpi$.
This equation is the Whittaker's Equation \cite{20} and
it has two basic solutions needed to construct the Green's
function (for ~$|\omega|\leq\mu$)
\begin{eqnarray}
\label{16}&&\tilde{\Psi}_{1}=M_{\beta,l+1/2}(2\varpi r)
=e^{-\varpi r}(2\varpi r)^{l+1}M(l+1-\beta,2l+2,2\varpi r), \\
\label{17}&&\tilde{\Psi}_{2}=W_{\beta,l+1/2}(2\varpi r)
=e^{-\varpi r}(2\varpi r)^{l+1}U(l+1-\beta,2l+2,2\varpi r),
\end{eqnarray}
where ~$M(a,b,z)$ and ~$U(a,b,z)$ are the two standard
solutions to the confluent hypergeometric equation \cite{20}.
~$U(a,b,z)$ is a many-valued function; i.e., there is a cut in
~$\tilde{\Psi}_{2}$.

From Eq. (\ref{9}), we find that the branch cut contribution
to the Green's function is
\begin{eqnarray}
\label{18}&&G^{c}(x,x';t)
=\frac{1}{(2\pi)^{2}}\sum_{l}\int_{-\mu}^{\mu}f(\varpi)
Y_{l}(\theta,a\omega)Y_{l}(\theta',a\omega)
e^{-i\omega t}d\omega,
\end{eqnarray}
where
\begin{eqnarray}
\label{19}&&\qquad f(\varpi)=\frac{\tilde{\Psi}_{1}(y',\varpi
e^{\pi i})\tilde{\Psi}_{2}(y,\varpi e^{\pi i})}
{W(\varpi e^{\pi i})}-\frac{\tilde{\Psi}_{1}(y',\varpi)
\tilde{\Psi}_{2}(y,\varpi)}{W(\varpi)}.
\end{eqnarray}
For simplicity we take ~$y>y'$. Of course,
this doesn't change the late-time behavior. One can
easily verify  that in the large ~$t$ limit the effective
contribution to the integral in Eq. (\ref{18}) arises from
~$|\omega|=0(\mu-1/t)$ or equivalently ~$\varpi=0(\sqrt{\mu/t})$.
This is due to the rapidly oscillating term ~$e^{-i\omega t}$
which leads to a mutual cancellation between the positive and
the negative parts of the integrand.

Using Eqs. (\ref{16}) and (\ref{17}), with the help of Eqs. 13.5.1 and
13.5.2 in Ref. \cite{20}, we have
\begin{eqnarray}
\label{20}&&\qquad\qquad W(\varpi e^{\pi i})=-W(\varpi)
=2(2l+1)\varpi\frac{\Gamma(2l+1)}{\Gamma(l+1-\beta)}.
\end{eqnarray}
In order to evaluate ~$G^{c}(x,x';t)$ using Eq. (\ref{18}), we only
need to study asymptotical form of ~$\tilde{\Psi}_{i}(y,\omega)
(i =1,2)$ both at the intermediate and very late times.

[{\bf Intermediate late-time tail}]:
It is by now well known that the intermediate asymptotic behavior
of a massive scalar field on the EMDA background is dominated
by flat spacetime effects \cite{5,21}. That is the tail in the range
~$M-D\ll r\ll t\ll (M-D)/[\mu (M-D)]^{2}$. In this time scale,
the frequency range ~$\varpi=0(\sqrt{\mu/t})$ implies ~$\beta\ll1$.
Notice that ~$\beta$ describes the effect of backscattering from
asymptotically far regions because it originates from the ~$1/r$ term
in the massive scalar field equation. If the relation ~$\beta\ll1$
is satisfied, the backscattering from asymptotically far
regions is negligible.

Using the following relations
\begin{eqnarray}
\label{28}&&W_{\beta,l+1/2}(2\varpi r)=\frac{\Gamma(-2l-1)}
{\Gamma(-l-\beta)}M_{\beta,l+1/2}(2\varpi r)+\frac{\Gamma(2l+1)}
{\Gamma(l+1-\beta)}M_{\beta,-(l+1/2)}(2\varpi r),\nonumber \\
\label{29}&&M_{\beta,l+1/2}(e^{\pi i}2\varpi r)=e^{(l+1)\pi i}
M_{-\beta,l+1/2}(2\varpi r),
\end{eqnarray}
and conditions $\beta\ll1$, $M(a,b,z)\simeq1$ as $z\rightarrow0$, Eq. (\ref{19})
can be expressed as
\begin{eqnarray}
\label{32}&&\qquad\qquad f(\varpi)=
\frac{2^{2l+2}\Gamma(-2l-1)\Gamma(l+1)}
{(2l+1)\Gamma(2l+1)\Gamma(-l)}
\left(yy'\right)^{l+1}\varpi^{2l+1}.
\end{eqnarray}
Substituting it into Eq. (\ref{18}) and performing the
integration, we obtain
\begin{eqnarray}
\label{21}&&G^{c}(x,x';t)
=\frac{1}{(2\pi)^{2}}\sum_{l}\frac{2^{3l+\frac{13}{2}}\Gamma(-2l-1)
\Gamma(l+\frac{3}{2})\Gamma(l+1)}
{(2l+1)\Gamma(2l+1)\Gamma(-l)}\mu^{l+\frac{1}{2}}
Y_{l}(\theta,a\mu)Y^{*}_{l}(\theta',a\mu)
\nonumber\\&&\qquad\qquad\qquad\times
(yy')^{l+1}t^{-(l+\frac{3}{2})}\cos[\mu t-(\frac{1}{2}l+\frac{3}{4})\pi].
\end{eqnarray}
Thus, the intermediate behavior is dominated by an
oscillatory inverse power-law decaying tail ~$t^{-(l
+3/2)}\sin(\mu t)$ for each multiple moment ~$l$.

[{\bf Asymptotically late-time tail}]:
In the above discussion we have used the approximation
of ~$\beta\ll1$, which only holds when ~$\mu t\ll1/[\mu(M-D)]^{2}$.
But at very late times ~$\mu t\gg1/[\mu(M-D)]^{2}$ the inverse power
law decay is replaced by another pattern of decay, which is slower than
any power law. This asymptotic tail behavior is caused by a
resonance backscattering due to spacetime
curvature \cite{14}. In this case ~$\beta$ is not negligibly
small, namely, the backscattering
from asymptotically far regions is important.

Now we have ~$\beta\gg1$. Using Eq. 13.5.13 of Ref. \cite{20}, we obtain
\begin{eqnarray}
\label{22}&&M_{\beta,\pm(l+1/2)}(2\varpi r)=
\Gamma(1\pm(2l+1))(2\varpi r)^{1/2}\beta^{\mp(l+1/2)}
J_{\pm(2l+1)}(\sqrt{8\beta\varpi r}),
\end{eqnarray}
where ~$J_{\pm(2l+1)}(z)$ is the Bessel function. Thus, we have
\begin{eqnarray}
\label{23}&&f(\varpi)=\frac{\Gamma(2l+2)\Gamma(-2l)}{(2l+1)}
\left(yy'\right)^{1/2}[J_{(2l+1)}(\sqrt{8\beta\varpi y'})
J_{-(2l+1)}(\sqrt{8\beta\varpi y})-I_{(2l+1)}(\sqrt{8\beta\varpi y'})
\nonumber\\&&\qquad\qquad\times
I_{-(2l+1)}(\sqrt{8\beta\varpi y})]+
\frac{[\Gamma(2l+2)]^{2}\Gamma(-2l-1)\Gamma(l+1-\beta)}{(2l+1)
\Gamma(2l+1)\Gamma(-l-\beta)}(yy')^{1/2}\beta^{-(2l+1)}
\nonumber\\&&\qquad\qquad\times
[J_{(2l+1)}(\sqrt{8\beta\varpi y'})J_{(2l+1)}(\sqrt{8\beta\varpi y})
+I_{(2l+1)}(\sqrt{8\beta\varpi y'})I_{(2l+1)}(\sqrt{8\beta\varpi y})],
\end{eqnarray}
where ~$I_{\pm(2l+1)}(z)$ is the modified Bessel functions.
Notice that ~$8\beta\varpi\simeq8(M-D)\mu^{2}$
is independent of ~$\omega$. Thus,
the late time tail arising from the first term will be ~$t^{-1}$. Now
let us discuss the second term (in the limit ~$\mu t\rightarrow\infty$
and ~$|\omega|\rightarrow\mu$)
\begin{eqnarray}
\label{24}&&G^{c}(x,x';t)
=\frac{1}{(2\pi)^{2}}\sum_{l}Q(l)\int_{-\mu}^{\mu}
\frac{\Gamma(l+1-\beta)}{\Gamma(-l-\beta)}\beta^{-(2l+1)}
e^{-i\omega t}d\omega,
\end{eqnarray}
where
\begin{eqnarray}
\label{25}&&Q(l)=\frac{[\Gamma(2l+2)]^{2}\Gamma(-2l-1)}
{(2l+1)\Gamma(2l+1)}(yy')^{1/2}
Y_{l}(\theta,a\mu)Y^{*}_{l}(\theta',a\mu)
\nonumber\\&&\qquad~~\times
\left[J_{(2l+1)}(\sqrt{8\beta\varpi y'})J_{(2l+1)}(\sqrt{8\beta\varpi y})
+I_{(2l+1)}(\sqrt{8\beta\varpi y'})I_{(2l+1)}(\sqrt{8\beta\varpi y})\right].
\end{eqnarray}
Since ~$\beta\gg1$, Eq. (\ref{24}) can be further approximated  to give
\begin{eqnarray}
\label{26}&&\qquad\qquad G^{c}(x,x';t)=\frac{1}{(2\pi)^{2}}\sum_{l}Q(l)
\int_{-\mu}^{\mu}
e^{i(2\pi\beta-\omega t)}e^{i\phi}d\omega.
\end{eqnarray}
Here the phase ~$\phi$ is defined by
~$e^{i\phi}=[1-e^{-i(2\pi\beta)}]/[1-e^{i(2\pi\beta)}]$.
The integral Eq. (\ref{26}) can be evaluated by the method of the saddle-point
integration \cite{14,15}. So we get
\begin{eqnarray}
\label{27}&&\qquad\qquad\qquad
G^{c}(x,x';t)\sim\frac{1}{(2\pi)^{2}}\sum_{l}Q(l)t^{-5/6}\sin(\mu t).
\end{eqnarray}
Obviously the very late-time tail behavior is dominated by an
oscillatory inverse power-law decaying tail ~$t^{-(5/6)}\sin(\mu t)$.

In summary, we have studied analytically both the intermediate and
asymptotically late-time behavior of massive scalar fields in the
background of a stationary axisymmetric EMDA dilaton black hole.
In the massive perturbations fields we find that if the field's
mass is small, namely ~$\mu(M-D)\ll1$, the intermediate tails
~$\Phi\sim t^{-(l+3/2)}\sin(\mu t)$ have been shown to dominate at
the intermediate late-time ~$\mu(M-D)\ll\mu t\ll1/[\mu(M-D)]^{2}$.
These oscillatory inverse power-law decaying tails originate from
the flat spacetime effects. Obviously, these tails depend not only
on the multiple moment ~$l$ but also on the field's mass $\mu$,
but they are independent of the rotational parameter $a$ of the
black hole. However, the intermediate late-time tails are not the
final pattern and a transition to the oscillatory asymptotically
tails ~$\Phi\sim t^{-5/6}\sin(\mu t)$ is to occur when ~$\mu
t\gg1/[\mu(M-D)]^{2}$. The origin of these tails may be attributed
to the resonance backscattering due to spacetime curvature. It is
interesting to mention that these behaviors depend only on ~$\mu$,
but they are independent of ~$l$ and ~$a$. Obviously, they are
qualitatively similar to those found in the Schwarzschild and
nearly extreme Reissner-Nordstr\"{o}m backgrounds. Our result
seems to suggest that the intermediate tails $\Phi\sim
t^{-(l+3/2)}\sin(\mu t)$ and the asymptotically tails ~$\Phi\sim
t^{-5/6} \sin(\mu t)$ may be a quite general feature for the
evolution of massive scalar fields in any  four dimensional
asymptotically flat rotating black hole backgrounds.

\begin{acknowledgments}This work was supported by the
National Natural Science Foundation of China under Grant No.
10275024; and the FANEDD under Grant No. 200317.
\end{acknowledgments}

\end{document}